\documentclass[aps,prl,twocolumn,superscriptaddress,footinbib,amsmath,amssymb]{revtex4}
\usepackage{dcolumn}
\usepackage{bm}
\usepackage{epstopdf} 
\usepackage{graphicx}
\usepackage{float}
\usepackage[english]{babel}

\begin{document}
\title{X-ray imaging non-destructively identifies functional 3D photonic nanostructures}

\author{D.A. Grishina}
\affiliation{COPS, MESA+ Institute for Nanotechnology, University of Twente, 
7500 AE Enschede, The Netherlands}
\affiliation{Present address: Thermo Fisher Scientific, Achtseweg Noord 5, 5651 GG Eindhoven, The Netherlands.}

\author{C.A.M. Harteveld}
\affiliation{COPS, MESA+ Institute for Nanotechnology, University of Twente, 
7500 AE Enschede, The Netherlands}

\author{A. Pacureanu}
\affiliation{European Synchrotron Radiation Facility (ESRF), B.P. 220, 
F-38043 Grenoble, France}

\author{D. Devashish}
\affiliation{COPS, MESA+ Institute for Nanotechnology, University of Twente, 
7500 AE Enschede, The Netherlands}
\affiliation{Present address: ASML Netherlands B.V., 5504 DR Veldhoven, The Netherlands.}

\author{A. Lagendijk}
\affiliation{COPS, MESA+ Institute for Nanotechnology, University of Twente, 
7500 AE Enschede, The Netherlands}

\author{P. Cloetens}
\affiliation{European Synchrotron Radiation Facility (ESRF), B.P. 220, 
F-38043 Grenoble, France}
\altaffiliation{Author for correspondence, email: cloetens@esrf.fr}

\author{W.L. Vos}
\affiliation{COPS, MESA+ Institute for Nanotechnology, University of Twente, 
7500 AE Enschede, The Netherlands}
\altaffiliation{Author for correspondence, email: w.l.vos@utwente.nl}

\date{August 3rd, 2018}

\begin{abstract}
\textbf{
To investigate the performance of three-dimensional (3D) nanostructures, it is vital to study \textit{in situ} their internal structure non-destructively.
Hence, we perform synchrotron X-ray holographic tomography on exemplary 3D silicon photonic band gap crystals without irreversible preparation steps. 
Here, we obtain real space 3D density distributions of whole crystals buried on 2 mm$^2$ beams with $20$ nanometer resolution. 
Our X-ray results identify why structures that look similar in scanning electron microscopy have vastly different nanophotonic functionality: 
One crystal with a broad photonic gap reveals 3D periodicity as designed ("Good"), a second structure without gap reveals a buried void ("Bad"), a third one without gap is shallow due to fabrication errors ("Ugly").
We conclude that X-ray tomography is a crucial tool to critically assess 3D functional nanostructures. 
}
\end{abstract}
\maketitle

\begin{figure}[h!]
\includegraphics[width=0.86\columnwidth]{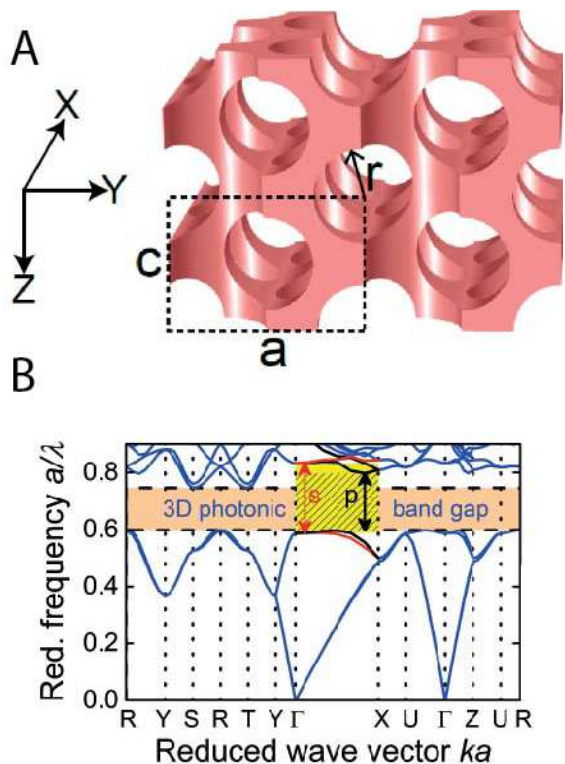}
\caption{ 
\textbf{Design of a 3D photonic crystal and its photonic functionality.}
(A) Cubic 3D inverse woodpile photonic crystals have a density distribution designed as two perpendicular 2D centered rectangular arrays (lattice parameters $a,c$; $a/c=\sqrt2$) of pores with radius $R$. 
Pores in the $X$-direction are aligned between rows of pores in the $Z$-direction~\cite{Ho1994SSC}. 
(B) Band diagram for an inverse woodpile crystal made from silicon \cite{Suppl} reveals a broad 3D photonic band gap between $a/\lambda = 0.60$ and $0.75$ (orange bar).
In the experimentally probed $\Gamma - X$ high symmetry direction (panel enlarged for clarity), the s-polarized stop gap (yellow) is broader than the p-polarized stop gap (black)~\cite{Devashish2017PRB}. 
\label{fig:design-and-bands}
}
\end{figure}

Three-dimensional (3D) nanostructures are drawing a fast-growing attention for their advanced functionalities in nanophotonics~\cite{Ergin2010Science,Soukoulis2011NP,Gansel2009Science,Wijnhoven1998Science,Noda2000Science,Joannopoulos2008book,Tandaechanurat2011NP}, photovoltaics~\cite{Bermel2007OE,Upping2011AM} and novel 3D integrated circuits and flash memories~\cite{Crippa2016book, IBM2012, Samsung2014}.
The functional properties of such nanostructures are fundamentally determined by their complex internal structure that consist of 3D arrangements of structural units such as spheres, rods, pores, or split-rings~\cite{Soukoulis2011NP,Gansel2009Science,Wijnhoven1998Science,Noda2000Science}.
As a representative example, we study here 3D periodic silicon photonic band gap crystals with a cubic diamond-like structure, see Fig.~\ref{fig:design-and-bands}A~\cite{Ho1994SSC,Maldovan2004NM}. 
These crystals are powerful tools to control the propagation and the emission of light on account of their broad complete 3D photonic band gap~\cite{Leistikow2011PRL,Devashish2017PRB}, see Fig.~\ref{fig:design-and-bands}B.
Inevitably, any fabricated nanostructure differs from its initial design, systematically in case of structural deformations~\cite{Fan1995JAP,Woldering2009JAP}, and statistically in case of size and positional disorder of the structural units~\cite{Hughes2005PRL,Koenderink2005PRB}. 
Consequently, the observed functionality differs from the expected one. 
It is therefore critical to assess the structure of a 3D nanomaterial and verify how well it matches the design. 

%
%

\begin{figure*}[tbp]
\[\includegraphics[width=2\columnwidth]{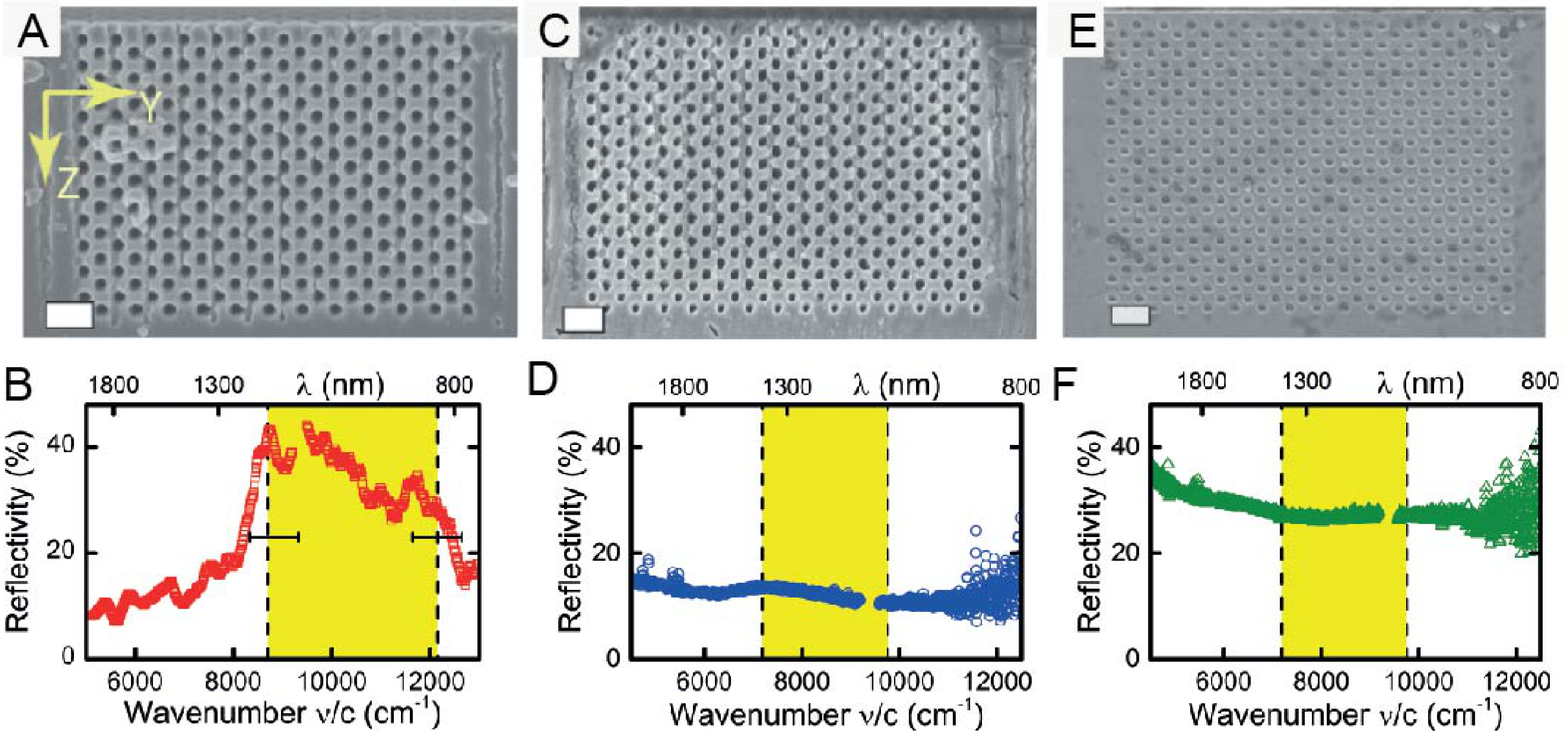}\]
\caption{
\textbf{SEM images and optical reflectivity spectra of three 3D photonic nanostructures.}
(A) SEM image of the external surface of a 3D inverse woodpile photonic crystal made from Si \cite{Suppl} whose measured reflectivity spectrum (B) reveals a broad photonic gap in agreement with theory (yellow hatched range). 
(C) SEM image of a 3D photonic crystal whose reflectivity spectrum (D) reveals a constant low reflectivity with no gap.
(E) SEM image of a 3D photonic crystal whose reflectivity spectrum (F) reveals a constant elevated reflectivity and no gap.
In (A,C,E) the scale bar is $1~\mu$m.
In (B) horizontal bars are estimated uncertainties in the stop gap width. 
\label{fig:SEM-and-function}
}
\end{figure*}

In nanotechnology, a fabricated sample is typically inspected by scanning electron microscopy (SEM)~\cite{Goldstein2003book}. 
A major limitation of SEM, however, is that only the external surface is viewed whereas the inner structure remains hidden. 
Indeed, Figure~\ref{fig:SEM-and-function} shows three 3D photonic-crystal nanostructures whose external surfaces look closely similar and closely match the design in Fig.~\ref{fig:design-and-bands}A. 
Remarkably, however, the corresponding nanophotonic functionality shown in Figure~\ref{fig:SEM-and-function} strongly differs: the crystal shown in panel A reveals a broad photonic gap as designed (panel B), whereas the other two structures reveal no gaps and instead a surprisingly constant reflectivity (panels D,F). 
To evaluate why a 3D nanostructure is functional or not, it is thus vital to determine the 3D structure non-destructively and \emph{in situ} with a technique that readily reveals local structural features. 

%
%

To visualize 3D nanostructures, SEM is supplemented with micro-machining or ion beam milling to cut away part of the structure~\cite{Goldstein2003book}.
Unfortunately, however, this approach is destructive, irreversible, and not \emph{in situ}.
While transmission electron microscopy allows for high-resolution 3D imaging, the required sample thickness of less than $1~\mu$m is insufficient for monolithic 3D photonic nanostructures~\cite{Jacobsen1998Book}.
X-ray techniques are well suited due to their high penetration and high resolution. 
While small-angle X-ray scattering is employed to study 3D nanoparticle arrays, it naturally operates in reciprocal space, making it hard to characterize local nano-sized features~\cite{Vos1997Langmuir,Shabalin2016PRL}.
In contrast, X-ray tomography yields the real space 3D distribution of the material density~\cite{Donoghue2006N, Sakdinawat2010NP}.
In traditional tomography, the contrast is provided by the sample absorption that is simply related to the brightness of the transmitted image called a radiograph~\cite{Pollak1953Chest}. 
Since silicon and many materials that prevail in nanotechnology and in complementary metal-–oxide-–semiconductor (CMOS) semiconductor industry absorb X-rays only weakly, however, advanced tomography methods are required.

Here, we obtain the relevant real space structural information directly from the optical \textit{phase change} of the X-ray beam that propagates through the sample.
The phase change is quantitatively retrieved from a set of radiographs taken at multiple sample-to-detector distances while rotating the sample~\cite{Suppl,Cloetens1999APL}.
Following a conventional tomographic reconstruction of the retrieved phase maps, the 3D electron density $\rho_{e} (X,Y,Z)$ is obtained in real space as a stack of equally spaced 2D slices in the plane normal to the sample's rotation axis. 
To achieve nanometer spatial resolution in a structure with thick (millimeter) substrates that do not need to be cut away, we employ X-ray holographic tomography~\cite{Mokso2007APL}. 
Its main features are that the X-ray beam is focused and that the sample is placed at a small distance $z_{s}$ downstream from the focus to collect magnified Fresnel diffraction patterns on the detector.

%
\begin{figure*}[tbp]
\[\includegraphics[width=2\columnwidth]{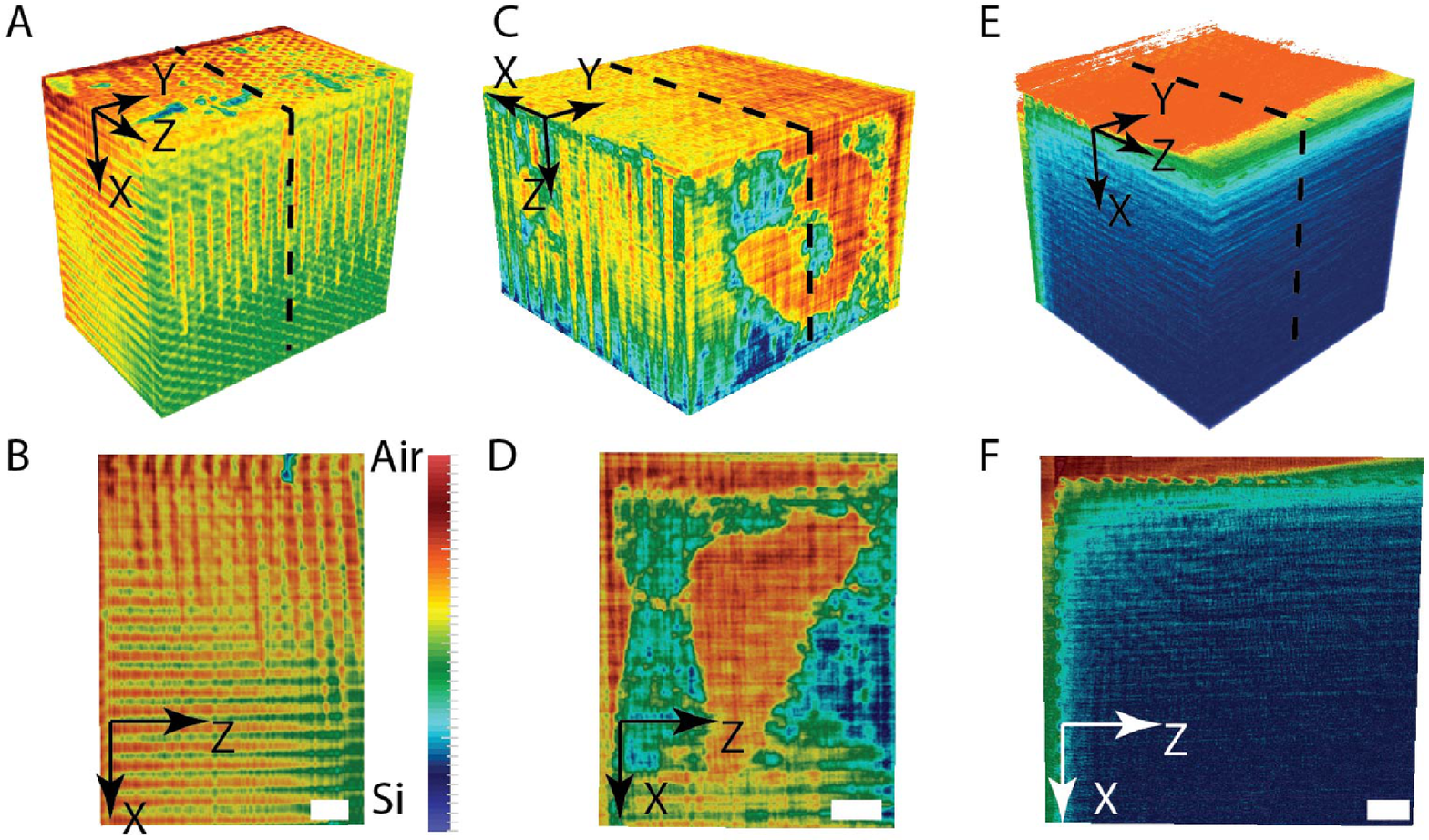}\]
\caption{\label{fig:tomo_main} 
\textbf{3D tomographic reconstructions of the three silicon nanostructures shown in the SEMs in Fig~\ref{fig:SEM-and-function}.} 
(A,C,E) Bird's-eye views of the reconstructed sample volume, $X,Y,Z$ axes are shown with each panel. 
(B,D,F) $XZ$ cross sections taken midway through each sample, a $1~\mu$m scale bar is shown in each slice. 
The sample in (A) is the same as in (B), in (C) as in (D), and in (E) as in (F). 
The common scale bar in panel (B) gives the electron density interpolated between silicon (blue) and air (red).
For animations of the data, see \cite{Suppl}. 
}
\end{figure*}
%
Figure~\ref{fig:tomo_main}A shows a bird's-eye view of the reconstructed sample volume of the 3D photonic crystal shown in Fig.~\ref{fig:SEM-and-function}A-B.
The $YZ$ top face shows the surface of the $X$-directed pores, similar to the SEM surface in Fig.~\ref{fig:SEM-and-function}A.
The alignment of the pores determines the 3D crystal structure and is a crucial step in the nanofabrication.
In practice, the alignment is controlled by the etch mask for each pore array and by the directionality of the etching processes~\cite{vandenBroek2012AFM}.
In the $XZ$ side face in Fig.~\ref{fig:tomo_main}A, pores are running in the $Z$ direction, whereas in the $XY$ front face, pores are running in the $X$ direction, matching the 3D design of the inverse woodpile structure (Fig.~\ref{fig:design-and-bands}A).
In the $XY$ front face several pores appear as if they are cut and start "from nowhere" in the middle, which is simply due to their running slightly obliquely to the $XY$ face, hence the top parts are not seen.

Figure~\ref{fig:tomo_main}B shows an $XZ$ cross-section midway through the 3D reconstructed volume that cuts through both arrays of pores and allows us to determine the maximum depths of both sets of pores. 
The $Z$-pores have a depth $D = 6280\pm 20$~nm and a radius $R = 183 \pm 10$~nm, corresponding to a state-of-the-art depth-to-diameter aspect ratio $17.15 \pm 0.04$, as expected from the deep reactive-ion etching settings~\cite{vandenBroek2012AFM,Wu2010JAP}. 
To date, the aspect ratio of pores deeply etched in silicon could only be assessed destructively and~\emph{ex-situ} by SEM inspection of ion-milled slices or cleaved cross sections~\cite{vandenBroek2012AFM,Wu2010JAP}. 
The deepest $X$-pores have an even greater depth of $9460\pm 20$~nm, corresponding to a high aspect ratio of $25.8 \pm 0.1$. 
Therefore, this is the first unequivocal observation that a second set of deep-etched pores runs even deeper than a first set.
Since the pore depth is a main limitation for a crystal's size, 3D nanostructures are thus significantly larger than expected before.
Clearly, Figure~\ref{fig:tomo_main}B already reveals buried structural features that are inaccessible to SEM or other nano-characterization methods (AFM, STM), thus illustrating the power of X-ray holographic tomography. 

In addition to characterizing functional nanostructures, X-ray tomography allows to identify several main deviations from design. 
Figure~\ref{fig:tomo_main}C-D show a bird's-eye view and a cross-section through a crystal whose external surface revealed usual crystalline features on a SEM image (\textit{cf.} Fig.~\ref{fig:SEM-and-function}C).
The tomographic reconstruction, however, reveals a buried internal void. 
The void is caused by well-known stiction (structural collapse)~\cite{Roman2010JPCM,Vos2011ECST}, as a result of strong capillary forces occurring on the nanoscale during the evaporation of liquid suspension of colloidal quantum dots that was infiltrated to study spontaneous emission~\cite{Leistikow2011PRL}. 
Thus from tomography we conclude that after these studies, the crystal lost its functionality as a photonic band gap device (see Fig.~\ref{fig:SEM-and-function}D. 
Figure \ref{fig:tomo_main}E-F show a bird's-eye view and a $ZY$ cross section of another sample whose external surface revealed usual crystalline features on a SEM image (see Fig.~\ref{fig:SEM-and-function}E). 
The tomographic reconstruction reveals a structure with pores that appear to be surprisingly shallow (about $70$ nm), as a result of inadvertent erroneous settings during the etching process. 
Thus tomography allows us to conclude that this peculiar structure has no band gap functionality to begin with (see Fig.~\ref{fig:SEM-and-function}F. 

In summary, we performed X-ray holographic tomography of 3D silicon photonic band gap crystals as a generic example of 3D nanomaterials. 
We obtain the 3D electron density and observe that the structural design is faithfully realized and leads to photonic functionality as expected.
In parallel, we uncover several buried structural deviations that help to identify the lack of photonic functionality of faulty structures.
We emphasize that the characterization method presented here is non-destructive, since a fabricated sample (up to $1 \times 2$ mm$^2$ cross-section) with buried nanostructures was mounted in the X-ray beam "as is" (see Methods), without the need for irreversible sample preparation steps. 
This feature contrasts to recent interesting work where samples had to be destructively milled to a much smaller size~\cite{Holler2017Nature}, or doped with heavy elements to obtain sufficient contrast~\cite{Chen2012AdvMat}.
We conclude that X-ray tomography is a powerful tool for structural characterization of any complex 3D functional nanostructure with arbitrary short or long-range order. 


\section{Acknowledgments}
This work was supported by the ''Stirring of light!'' program of the  "Nederlandse Organisatie voor Wetenschappelijk Onderzoek" (NWO), by the NWO-domain "Toegepaste en Technische Wetenschappen" (TTW) nr. 11985, the Shell-NWO/FOM programme ``Computational Sciences for Energy Research" (CSER), the MESA$^{+}$ Institute for Nanotechnology (Applied Nanophotonics (ANP)), and by the 2014 Descartes-Huygens Prize of the French Academy of Sciences to W.L.V. 
We thank ESRF for granting beamtime through experiments HS-2520 and CH-5092.

We thank Leon Woldering, Hannie van den Broek, Willem Tjerkstra, Simon Huisman, Rajesh Nair, Elena Pavlenko, Mehdi Aas, and the MESA+ Nanolab and the ESRF staff for help, and Arie den Boef (ASML), Jean-Michel G\'erard (Grenoble), Hans Hilgenkamp, Detlef Lohse, Allard Mosk (Utrecht), Pepijn Pinkse, and Hasan Yilmaz (Yale) for fruitful discussions.

W.L.V. and P.C. conceived the main idea, D.A.G. and C.H. prepared the samples, D.A.G., C.H., A.P., P.C., W.L.V. conducted synchtrotron experiments. 
D.A.G. and P.C. analyzed the X-ray data, A.P. made the animations. 
D.A.G., C.H. and W.L.V. conducted optical experiments, D.D. performed theoretical band structure calculations. 
D.A.G. and W.L.V. wrote the manuscript, with input from all authors. 
A.L., P.C., and W.L.V. supervised the research. 
All authors contributed to the analysis and discussion of the results. 
All data needed to evaluate the conclusions are present in the paper and/or the supplementary materials. 
Additional data related to this paper may be requested from the authors. 

\newpage

\section{Supplementary materials}

\subsection{Materials and methods}

\subsubsection{3D photonic crystal fabrication}
The fabrication process of our 3D photonic band gap crystals is described in detail in Refs.~\cite{vandenBroek2012AFM,Grishina2015NT}. 
In brief, a hard mask is defined on a silicon wafer with the centered rectangular array of apertures, with a pore radius $R/a = 0.245$ that gives the broadest possible band gap~\cite{Woldering2009JAP}.
Deep reactive ion etching of the first set of deep pores results in a wafer filled with a large 2D array of deep pores running in the $Z$-direction~\cite{vandenBroek2012AFM}.  
Next, the wafer is cleaved and polished, and the second hard mask is carefully aligned and defined in a $10 \times 10~\mu m^2$ area on the side face of the wafer. 
By etching the second set of pores in the $X$-direction, the 3D structure is obtained in the volume where both sets of pores overlap, see Figure~\ref{fig:tomo_main} and the movie animations M1-M2. 
Finally, the hard mask is removed. 
3D photonic crystals shown in Figure~\ref{fig:tomo_main}A-D are fabricated in the above mentioned way and are buried on chips with cross-section up to $1 \times 2$~mm$^2$. 

In our second generation of photonic crystals, the etch mask is deposited in a single step on both faces of a wafer edge~\cite{Grishina2015NT}, followed by deep reactive ion etching of two perpendicular arrays of pores. 
The 3D photonic crystal shown in Figure~\ref{fig:tomo_main}E-F is an example of a second generation photonic crystal fabricated with the single-step etch mask. 
While this particular sample turned out to be not successful, this fabrication route has yielded many other samples that had the intended 3D structure, as confirmed by X-ray tomography. 

\begin{figure}[h!]
\includegraphics[width=0.8\columnwidth]{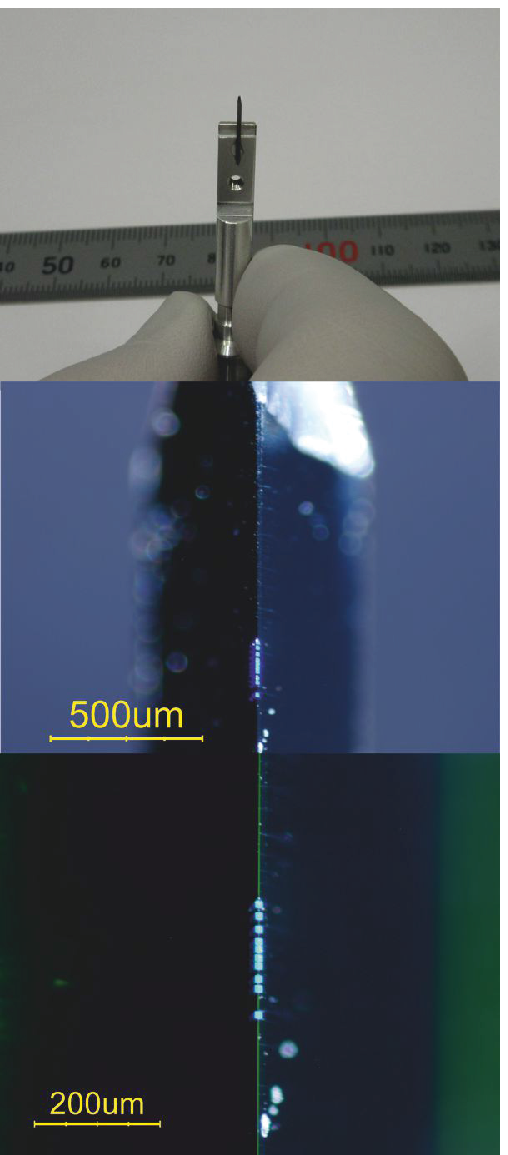}
\caption*{\label{fig:photo}
\textbf{Figure S1. 
Photographs of a typical sample studied by X-ray tomography.} 
Top: A silicon beam with photonic crystal structures is mounted in a holder for the X-ray tomography scans. 
Center: Zoom-in of the top part of the Si beam, with a vertical row of 3D photonic crystal structures on the edge of the beam.
In the defocused background, the edges of the beam's inclined surfaces are visible. 
Bottom: Further zoom-in reveals ten 3D photonic crystal structures that display a blueish iridescence due to their periodic surface structure. 
The edge of the beam appears as the vertical green line of scattered light.
}
\end{figure}

Figure~\ref{fig:photo} (S1) shows photographs of a sample as it is studied in the X-ray tomography instrument. 
The silicon beam measures $0.5 \times 0.5 \times 10$ mm$^3$ and is shown after preparation in the MESA+ Nanolab, and mounted for X-ray tomography scans in the ESRF. 
We emphasize that there is no need to cut a specific area out of the sample using focused-ion beam (FIB) milling, as opposed to other imaging techniques that require small sample volumes, such as X-ray ptychography, transmission electron microscopy (TEM), FIB-SEM, and so forth. 
We have successfully mounted several samples characterized by X-ray holographic tomography at ESRF in optical setups in Twente without further modifications, and even in the same sample holder.

A feature in Figure~\ref{fig:tomo_main}B is that the pores running in the $X$-direction are not exactly perpendicular to the pores in the $Z$-direction. 
It appears that the angle between the pore arrays deviates from the $90^{o}$ design by $2^{o}$; from results on several crystals we find deviations between $0^{o}$ and $6^{o}$. 
We attribute the slight variations to the limited precision in placing a sample in the etching chamber. 
From separate plane-wave band structure calculations we find that the 3D photonic band gap is robust to the slight shear of the cubic crystal structure~\cite{Woldering2009JAP}. 

\subsubsection{Synchrotron X-ray holographic tomography}

\begin{figure}[h!]
\includegraphics[width=1.0\columnwidth]{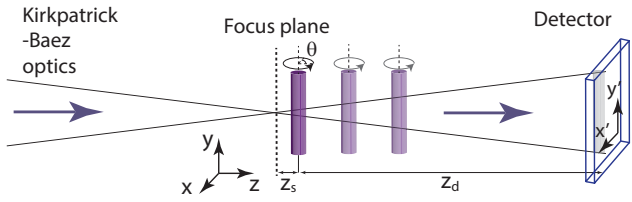}
\caption{\label{fig:tomo-schematic}
\textbf{(Figure S2.) 
Schematic of the holotomographic setup at ESRF.} 
The incident X-ray beam is focused using Kirkpatrick-Baez (KB) optics into a $23\times37$~nm$^2$ focus. 
The sample is placed at a small distance $z_s$ downstream from the focus, while the detector is placed at the distance $z_d$.
Tomographic scans are recorded for multiple sample-to-detector distances.
}
\end{figure}

Holographic tomography experiments were performed at the European Synchrotron Radiation Facility (ESRF), on the nano-imaging beamline ID16A-NI~\cite{Dasilva2017Optica}.
The X-ray beam with $17$ keV photon energy propagates in the $z$-direction and is focused with multilayer coated Kirkpatrick-Baez optics to a $23\times37$~nm$^2$ focus. 
The sample is placed at a small distance $z_s$ downstream from the focus, while the detector is placed at the distance $z_d$ downstream from the sample, see Figure~\ref{fig:tomo-schematic}. 
The image recorded in the detector plane is an in-line Gabor hologram or Fresnel diffraction pattern~\cite{Gabor1948Nature}. 
Due to the focusing, the sample is illuminated with a spherical wave, unlike the plane-wave illumination in traditional tomography. 
According to the Fresnel scaling theorem, the spherical wave illumination gives rise  to an effective propagation distance $D$ and a magnification $M$ given by~\cite{Pogany1997RSI, Paganin2006book}
\begin{equation}\label{eq:magnification}
{D=\frac{z_sz_d}{z_s+z_d},~M=\frac{z_s+z_d}{z_s}}.
\end{equation}

Varying the focus-to-sample distance $z_s$ allows us to vary the magnification of the diffraction patterns. 
It also strongly modifies the Fresnel diffraction pattern recorded on the detector through the effective propagation distance. 
For a phase periodic object such as our photonic band gap crystals, the Talbot effect results in zero contrast for certain spatial frequencies at the characteristic Talbot distances~\cite{Goodman2005Book}. 
To obtain non-zero contrast at all spatial frequencies, data are taken at four distances $z_s$. The first distance was chosen to obtain a desired pixel size, either 10 or 20 nm.
At each distance $z_s$, $1500$ images were recorded with $0.3~$s exposure time while rotating the sample from $\theta = 0^{o}$ to $180^{o}$ around the $Y$-axis of the crystal (see Fig.~\ref{fig:SEM-and-function}) . 
For the tomographic scans, the axis of rotation was aligned to be a few $\mu m$ deep inside the silicon. 

The data processing is a two-step procedure consisting of a phase retrieval step followed by a tomographic reconstruction. 
The phase retrieval aims at retrieving the amplitude $A(x,y)$ and phase $\phi(x,y)$ of the wave exiting the sample $u_0 = A(x,y) e^{i \phi(x,y)}$ and that are given by
\begin{equation}\label{eq:amplitude}
A(x,y)=exp(-\frac{2 \pi}{\lambda}\int \beta(x,y,z) dz)
\end{equation}
and
\begin{equation}\label{eq:phase}
\phi(x,y)=-\frac{2 \pi}{\lambda}\int \delta(x,y,z) dz .
\end{equation}
with $\lambda$ the X-ray wavelength. 
The amplitude and phase are thus the projection of respectively the absorption index $\beta$ and the refractive index decrement $\delta$ that determine the complex refractive index for hard X-rays
\begin{equation}\label{eq:refractive-index}
n(x,y,z) = 1 - \delta(x,y,z) + i \beta(x,y,z) .
\end{equation}
Prior to phase retrieval, all sets of radiographs are scaled to the same magnification and mutually aligned. 
We determine a first estimate of the amplitude and phase using the approach proposed by Paganin \textit{et al.}~\cite{Paganin2002JM}, and extended to multiple distances. 
Here, we assume a homogenous ratio $\delta / \beta = 174$ for silicon at an X-ray energy of $17~$keV. 
This first estimate provides only a blurred version of the phase map. 
The map is recursively improved using $15$ iterations of a non-linear least squares optimization. 
The phase retrieval was carried out with ESRF in-house software using the GNU Octave programming environment (www.octave.org) and the public domain image analysis program ImageJ (see http://rsbweb.nih.gov/ij). 

Secondly, a standard tomographic reconstruction~\cite{Hounsfield1973BJR} based on the filtered back-projection algorithm~\cite{Herman2009Book} and implemented in the ESRF software PyHST2~\cite{Mirone2014NIMB} allows us to obtain the distribution of the refractive index decrement $\delta(x,y,z)$. 
Since the X-ray energy of $17$ keV is far above any absorption edge of the materials under investigation, we obtain the electron density distribution $\rho_e (x,y,z)$ from the well-known expression~\cite{Guinier1963Book}
\begin{equation}\label{eq:electron-density}
\rho_e (x,y,z)=\frac{2\pi}{r_e \lambda^2}\delta(x,y,z)
\end{equation}
with $r_e$ the classical electron radius.
The resulting structure was rendered with open-source software ParaView (see www.paraview.org). 

\subsubsection{Nanophotonic experiments}
To assess the basic functionality of the photonic crystals, we performed optical reflectivity to probe the designed photonic gaps. 
Optical reflectivity was measured using a home-built microscope setup that employs reflective optics and operates in the near infrared range (at wavelengths beyond $800$~nm, see Ref.~\cite{Huisman2011PRB}.
Main components are a supercontinuum white light source (Fianium), a Fourier-transform interferometer (Biorad FTS-6000) that operates with $8$ cm$^{-1}$ spectral resolution, and a reflecting objective $NA=0.65$ to focus the beam to a few microns inside the photonic crystal domains over the large required range of frequencies. 
Signals were collected in the near infrared spectral range between about $4000$~cm$^{-1}$ and more than $12500$~cm$^{-1}$ (corresponding to wavelengths between $2500$ and $800$~nm).
Reflectivity was calibrated by taking the ratio of a sample spectrum with the spectrum measured on a clean gold mirror.
In the spectra in Figs.~\ref{fig:SEM-and-function}B,D,F a narrow range near $9300$ cm$^{-1}$ is omitted since it is disturbed by the pump laser of the supercontinuum source.
The maximum reflectivity in Fig.~\ref{fig:SEM-and-function}B is limited to $45 \%$ due to several different reasons: the white light focus has a diameter ($4~\mu$m) comparable to the extent over which the crystal is homogeneous; near the crystal-air interface the pores reveal roughness as a result of the etching process (so-called "scallops"~\cite{Wu2010JAP,vandenBroek2012AFM}).

\subsubsection{Photonic band structure calculations}
Photonic band structures were calculated with the plane-wave expansion method, using the MIT photonic bands (MPB) code~\cite{Johnson2001OE}.
Silicon was modeled with a dielectric function $\epsilon = 12.1$, and the unit cell was discretized with a high spatial resolution $(\Delta X, \Delta Y, \Delta Z) = (c/140, a/200, c/140)$. 
The band structures are represented as reduced frequency $a/\lambda$ versus reduced wave vector $k a$ between high-symmetry points (R, Y, S, R, T, Y, $\Gamma$, X, U, $\Gamma$, Z, U, R) in the 1st Brillouin zone of the convenient tetragonal representation of the unit cell~\cite{Joannopoulos2008book}. 
The polarization states of the photonic bands that bound the $\Gamma-X$ stop gap (or symmetrically equivalently the $\Gamma-Z$ gap) (see Figure~\ref{fig:design-and-bands}B) were assigned based on our recent computational study~\cite{Devashish2017PRB}.
The photonic bands in Figure~\ref{fig:design-and-bands}B and the gap in Figure~\ref{fig:SEM-and-function}B were calculated for a relative pore radius $R/a=0.267$ as borne out of the X-ray tomography data, and the gaps in Figures~\ref{fig:SEM-and-function}D,F were calculated for the designed pore radius $R/a=0.245$. 

\subsection{Multimedia files}
In view of the richness of the 3D reconstructions, we provide three movie animations of rotations about and cross-sections through the 3D data volumes, in AVI format. 

\begin{enumerate}

\item Animation 'theGood1.avi': 
Color animation of a rotation of the successfully etched crystal shown in Fig.~3A-B about the vertical axis in Fig.3~A. 
The animation shows the reconstructed sample volume from all viewing directions. 
Still at $0:01$: 
\begin{figure}[h!]
\begin{center}
\end{center}
\end{figure}

\item Animation 'theGood2.avi': 
Black and white animation of YZ cross-sections scanned in the X-direction through the volume of the successfully etched crystal shown in Fig.~3A-B. 
The scan starts from the bottom in the substrate, scans parallel to X-oriented pores to show Z-oriented pores, and finishes in air above the crystal. 
Still at $0:11$: 
\begin{figure}[h!]
\begin{center}
\end{center}
\end{figure}

\item Animation 'theBad.avi': 
Black and white animation of YZ cross-sections scanned in the X-direction through the volume of the crystal with a void shown in Fig.~3C-D. 
The scan starts from the bottom in the substrate, scans parallel to X-oriented pores while traversing the stiction-induced void (where the surrounding 2D photonic crystal pores are still apparent left and right), and finishes in air above the crystal surface.
Still at $0:09$: 
\begin{figure}[h!]
\begin{center}
\end{center}
\end{figure}


\end{enumerate}

\noindent
Since the unetched ("the Ugly") sample (Fig.~3E-F) reveals mostly cross-sections through bulk silicon, we have decided against an animation for this sample. 


\end{document}